\documentstyle[11pt]{article}
\newskip\humongous \humongous=0pt plus 1000pt minus 1000pt

\newif\ifdtup

\def\Tr{\mathop{\rm Tr}}

\def\beq{\begin{equation}}
\def\eeq{\end{equation}}

\def\beqn{\begin{eqnarray}}
\def\eeqn{\end{eqnarray}}
\relax

\def\dotx{\dotx{\dot\overline{x}}}

\relax

\catcode`\@=11

\@addtoreset{equation}{section}
\def\theequation{\arabic{section}.\arabic{equation}}

\def\@normalsize{\@setsize\normalsize{15pt}\xiipt\@xiipt
\abovedisplayskip 14pt plus3pt minus3pt%
\belowdisplayskip \abovedisplayskip
\abovedisplayshortskip  \z@ plus3pt%
\belowdisplayshortskip  7pt plus3.5pt minus0pt}

\def\small{\@setsize\small{13.6pt}\xipt\@xipt
\abovedisplayskip 13pt plus3pt minus3pt%
\belowdisplayskip \abovedisplayskip
\abovedisplayshortskip  \z@ plus3pt%
\belowdisplayshortskip  7pt plus3.5pt minus0pt
\def\@listi{\parsep 4.5pt plus 2pt minus 1pt
            \itemsep \parsep
            \topsep 9pt plus 3pt minus 3pt}}

\@twosidetrue
\relax

\catcode`@=12

\catcode`\@=11

\def\section{\@startsection{section}{1}{\z@}{3.5ex plus 1ex minus
   .2ex}{2.3ex plus .2ex}{\large\bf}}

\def\thesection{\Roman{section}.}

\def\appendix{\setcounter{section}{0}
	\def\thesection{APPENDIX \Alph{section}:}
	\def\theequation{\Alph{section}.\arabic{equation}}}
\def\FERMIPUB{}
\def\FERMILABPub#1{\def\FERMIPUB{#1}}
\def\ps@headings{\def\@oddfoot{}\def\@evenfoot{}
\def\@oddhead{\hbox{}\hfill
	\makebox[.5\textwidth]{\raggedright\ignorespaces --\thepage{}--
	\hfill {\rm FERMILAB--Pub--\FERMIPUB}}}
\def\@evenhead{\@oddhead}
\def\subsectionmark##1{\markboth{##1}{}}
}

\ps@headings

\catcode`\@=12

\relax

\def\figcap{\section*{Figure Captions\markboth
	{FIGURECAPTIONS}{FIGURECAPTIONS}}\list
	{Fig. \arabic{enumi}:\hfill}{\settowidth\labelwidth{Fig. 999:}
	\leftmargin\labelwidth
	\advance\leftmargin\labelsep\usecounter{enumi}}}
 \relax
\def\tablecap{\section*{Table Captions\markboth
	{TABLECAPTIONS}{TABLECAPTIONS}}\list
	{Table \arabic{enumi}:\hfill}{\settowidth\labelwidth{Table 999:}
	\leftmargin\labelwidth
	\advance\leftmargin\labelsep\usecounter{enumi}}}
 \relax
\def\reflist{\section*{References\markboth{REFLIST}{REFLIST}}\list
	{[\arabic{enumi}]\hfill}{\settowidth\labelwidth{[999]}
	\leftmargin\labelwidth
	\advance\leftmargin\labelsep\usecounter{enumi}}}
 \relax

\catcode`\@=11

\def\FERMIPUB{}
\def\FERMILABPub#1{\def\FERMIPUB{#1}}
\def\ps@headings{\def\@oddfoot{}\def\@evenfoot{}
\def\@oddhead{\hbox{}\hfill
	\makebox[.5\textwidth]{\raggedright\ignorespaces --\thepage{}--
	\hfill {\rm FERMILAB--Pub--\FERMIPUB}}}
\def\@evenhead{\@oddhead}
\def\subsectionmark##1{\markboth{##1}{}}
}

\ps@headings

\relax

\FERMILABPub{92/226--T}

\newcommand{\tint}{$\tau_{int}$}
\newcommand{\Arg}[1]{{\rm Arg\left\{#1\right\}}}
\newcommand{\Real}[1]{{\rm Re\left\{#1\right\}}}
\newcommand{\half}{{\scriptstyle{1\over 2}}}
\newcommand{\oneahalf}{{\scriptstyle{3\over 2}}}
\newcommand{\onethird}{{\scriptstyle{1\over 3}}}

\FERMILABPub{92/226--T}
\begin{document}
\title{}
\author{}
\maketitle
\begin{flushright}
FERMILAB--PUB--92/226--T \\
\end{flushright}
\begin{center}
{\bf Embedding $Z(3)$ in $SU(3)$}\\[.25in]
{\bf E. Onofri\\
{\it Theory Department}\\
{\it Fermi National Accelerator Laboratory} \\
{\it P.O.Box 500, Batavia, IL 60510}}
\footnote{\rm Permanent address:
Dipartimento di Fisica,
Universit\`a di Parma,
and I.N.F.N., Gr.Coll. di Parma,
43100 Parma, Italy,
E-mail:\sl{onofri@vaxpr.pr.infn.it}}\\
\date{August 15, 1992}

\begin{abstract}
We consider the design of  a non-local
MonteCarlo algorithm for $SU(3)$ lattice systems
according to the idea of {\em embedding} the degrees of
freedom corresponding to the center of the group $Z(3)$.
As a crucial ingredient to reach this goal, we present a  practical
implementation of a cluster algorithm for
$Z(3)$ systems with general random pair interaction.
\end{abstract}
\end{center}
\newpage
\section{Introduction}
\noindent
Developing non-local upgrade mechanisms for $SU(N)$ lattice
models proved to be a rather hard task, because of the
impossibility to carry on a straightforward generalization
of Wolff's embedding idea.\cite{CEPS}
There are apparently no such algorithms
available for $SU(N)$ lattice systems, at present.
The relatively small effort devoted to this goal has
probably to do with the fact that the dynamics of gauge fields
takes only a small part of the total computer time, and
the feeling is that there is little to be gained in
optimizing it. Still, large statistical errors on the
gluon dynamics may result when large lattices will
be used to explore more deeply the scaling region.
 We take the point of view that sometime it will
pay to have an efficient algorithm for the pure gauge
dynamics and we want to argue
in this paper that  a strategy similar to
Brower-Tamayo's \cite{Brower}
embedding may possibly work, as  suggested by Wolff \cite{UW}. We
present some preliminary steps to be taken to
develop such an algorithm, even if their true value
will be appreciated (hopefully) only after these ideas
will be implemented in a realistic calculation.
In order to simplify the presentation, we  explicitly  cover
the {\sl N=3\/}  two-dimensional sigma-model,
but there seem to be no obstructions to generalize the idea
to the case of four-dimensional QCD at finite temperature,
which is the kind of real application we have in mind. According
to ideas of Polyakov going back to the 70's, it is precisely
the center of the group which governs the deconfining transition,
hence the $Z(3)$ degrees of freedom should be responsible for
a critical slowing down near the critical temperature.

In sect.\ref{sec:embed} we describe the simple embedding we have
in mind for $SU(3)$; in sect.\ref{sec:algo} we derive the algorithm
and we give a
few details on its implementation while in sect.\ref{sec:numer} we
present some preliminary numerical results.

\section{$Z(3)$ embedding}\label{sec:embed}
Embedding $Z(3)$ degrees of freedom into a $SU(3)$ model, like
the non-linear sigma model,
can be accompished by assigning a cross section of the fibre bundle
$ Z(3)\to SU(3) \to SU(3)/Z(3) $. For instance such a section may be
chosen as
\[ 	V \in SU(3) \to \zeta \cdot U \cases{ \zeta \in Z(3)&\cr
					    U \in SU(3)/Z(3)&\cr}
\]
\begin{equation}\label{eq:coset}
 | \Arg{\Tr{U}} | < \pi/3
\end{equation}
Since the bundle is non-trivial, the section is only defined up
to a set of measure zero, but this is fine for our purposes.
The action transforms as follows
(here and in the following $<ij>$
implies that $i$ and $j$ are nearest neighbours):
\begin{eqnarray*}
S & = & \sum_{<ij>} \onethird \Real{\Tr{V_i^\dagger V_j}}  \\
  & = & \sum_{<ij>} \Real{ \bar\zeta_i \zeta_j
   \; \onethird \Tr{U_i^\dagger U_j}}
\end{eqnarray*}
\noindent
and it appears as a $Z(3)$ model in the variables $\zeta$ with
random couplings. In the {\sl gauge\/} case, we would end up with
a $Z(3)$ lattice gauge theory with random coupling.
Near the continuum limit the matrix $U_i^\dagger U_{j}$ fluctuates
around the identity, and the system is effectively ferromagnetic.

To exploit this embedding to beat critical slowing down at the
continuum limit one needs an efficient algorithm for the
$Z(3)$ degrees of freedom. While a cluster algorithm
for $Z(3)$ spin systems is well-known in the real-constant-coupling
case (a special case of {\sl q-}Potts model with {\sl q=3\/}), the random
coupling case requires some modification. A general scheme
for local pair interaction has been devised by Niedermayer
\cite{Nieder}
and it covers the present situation. We are going to derive
a cluster algorithm along a slightly different line of reasoning,
but the outcome will be the same. Instead
of concentrating on transition probabilities satisfying
detailed balance, we shall introduce an equivalent ensemble
with bond variables as new degrees of freedom, like one
does in the Ising model. There are two ways to define the
action in terms of bond variables and these correspond to
two special cases in Niedermayer's approach.
The resulting cluster dynamics is non trivial in the first case
while clusters are statistically independent
in the second case.

\section{The algorithm}\label{sec:algo}

Let us set the notation: $ \zeta \in Z(3) $ is a generic
spin variable, its values ranging on the set
$ (1, \omega, \bar\omega)$, where $ \omega = \exp(2\pi i/3) $ .
The action we consider is the following
\begin{equation} \label{eq:action}
S = \sum_{<ij>} \Real{ \bar\zeta_i \Omega_{ij} \zeta_j }
\end{equation}
where $i,j$ range over pairs of next-neighbours.

The coupling constants $\Omega_{ij}$ are subject to a limitation
which is not essential for the algorithm to work, but
will simplify the presentation, namely we assume that
\begin{equation} \label{eq:cond}
| \Arg{\Omega_{ij}} | <  2 \pi/3
\end{equation}
for all $(i,j)$, a condition which is likely to be met near
the continuum limit.
Now let us write the partition function in the following way:
$$ Z = \prod_{<ij>} Z_{ij} $$
\begin{eqnarray*}
Z_{ij} & = &\exp(\beta\Real{\Omega_{ij}}) \delta_{\zeta_i,\zeta_j} \\
 & & \mbox{}+ \exp(\beta\Real{\Omega_{ij}\bar\omega})
	\delta_{\zeta_i,\omega\zeta_j}\\
 & & \mbox{}+ \exp(\beta\Real{\Omega_{ij}\omega})
	\delta_{\zeta_i,\bar\omega\zeta_j}
\end{eqnarray*}
Ignoring a common factor $ \prod \exp(\beta\Real{\Omega_{ij}}) $,
we can rewrite $Z$ as follows:
\begin{equation}
Z =  \prod_{<ij>} \left(\delta_{\zeta_i,\zeta_j} +
  W_{ij}\,(\bar\omega) \delta_{\zeta_i,\omega\zeta_j} +
  W_{ij}\,(\omega) \delta_{\zeta_i,\bar\omega\zeta_j}\right)
\end{equation}
where
\[
W_{ij}(z) = \exp(\beta\Real{(z-1)\,\Omega_{ij}})
\]
On each link $ij$ let us define
\[
 W_{ij}^< = {\rm min} \left( W_{ij}(\omega), W_{ij}(\bar\omega)\right)
\]
\begin{equation}
W_{ij}^> = {\rm max} \left( W_{ij}(\omega), W_{ij}(\bar\omega)\right)
\end{equation}
and
\begin{equation}
\omega_{ij}^\flat = \cases{ \omega & if $W^> = W(\omega)$\cr
			    \bar\omega & otherwise }
\end{equation}

Under the condition above on the phase of $\Omega$ (Eq.\ref{eq:cond}),
 it turns out that
$W^< < 1$; as a consequence, if we  rewrite $Z_{ij}$ as follows:
\begin{equation}
Z_{ij} = \delta_{\zeta_i, \zeta_j} + W_{ij}^< \,
	(1 - \delta_{\zeta_i, \zeta_j})
+ (W_{ij}^> - W_{ij}^<) \delta_{\zeta_i, \omega_{ij}^\flat\zeta_j}.
\end{equation}
 we can introduce bond variables $n_{ij}$ as in Swendsen-Wang's
algorithm, to get either

\begin{equation}
Z = \sum_{\zeta, n} \prod_{<ij>}
	\left\{
	\delta_{n_{ij},1} \delta_{\zeta_i,\zeta_j} (1 - W_{ij}^<) +
        \delta_{n_{ij},0} [ W_{ij}^< + (W_{ij}^> - W_{ij}^<)\,
	\delta_{\zeta_i,\omega_{ij}^\flat\zeta_j} ]
	\right\}
\end{equation}
or
\begin{equation}\label{eq:tredi}
Z = \sum_{\zeta, n} \prod_{<ij>}
 \left\{
 \delta_{n_{ij},1} [\delta_{\zeta_i,\zeta_j}\, (1 - W_{ij}^<) +
 \delta_{\zeta_i,\omega_{ij}^\flat\zeta_j} \, (W_{ij}^> - W_{ij}^<)] +
 \delta_{n_{ij},0} W_{ij}^< \right\}
\end{equation}

As a first option we then have the following updating algorithm:
{\em first step}, the
bonds are switched  on with probability $(1 - W_{ij}^<)$, provided
$\zeta_i = \zeta_j$ (this first step is identical to Swendsen-Wang's).
The {\em second step} is given by
the spin update, which is governed by the term $\delta_{n_{ij},1}$,
 forcing the same spin value on each connected cluster. These
cluster spin values are chosen according to a distribution
$Z_{clust}$ which
takes into account cluster interaction  which occurs at the
boundaries. The effective action for the cluster dynamics can be
extracted from the term involving $\delta_{n_{ij},0}$. Let ${\cal C}$
denote a generic cluster, $\zeta_{{\cal C}}$ its spin; then we have
\[
Z_{clust} = \prod_{{\cal C, C'}} W(\zeta_{{\cal C}}, \zeta_{{\cal C'}})
\]
where
\[
	W(\zeta_{{\cal C}}, \zeta_{{\cal C'}}) =
	\prod_{ <ij>}^{
	i \in \partial{\cal C}, j \in \partial{\cal C'}}
	\left(W_{ij}^< + (W_{ij}^> - W_{ij}^<)\,
	\delta_{\zeta_{{\cal C}}, \omega_{ij}^\flat\zeta_{{\cal C'}}}
	\right)
\]
The clusters' interaction dictated by this formula can be now
treated as an ordinary lattice system, {\it e.g.} by a
heat bath method. What we have realized here is a transformation
from the original $Z(3)$ lattice system to another $Z(3)$ system
indexed by clusters.  The implementation which
we have already tested (see next section)  is suggested by Wolff's
{\it single cluster} algorithm. The assumption underlying such a
choice is that in some equilibrium regime the existing spin
values on the boundary of the cluster can be used to evaluate
the probability distribution even without growing the other
clusters.

As for the second option, starting from Eq.\ref{eq:tredi}
the bonds are switched on with probability
\[
	\cases{1-W^< & if $\zeta_i=\zeta_j$\cr
	  1- {W^< \over W^>} & if $\zeta_i=\omega^\flat_{ij}\zeta_j$\cr
	  0 & 	otherwise\cr}
\]
 There is no interaction between clusters
(which may be read from the fact that the coefficient $\delta_{n,0}$
is $\zeta$-independent), but the interaction among spins within
a cluster is non trivial. Instead of trying to thermalize
the system on each cluster, we can just apply a global rotation
chosen independently at random on each cluster. Niedermayer
showed that this move satisfies detailed balance and ergodicity
is obviously satisfied since clusters with just one site are
possible.

The setup outlined here would not work in case Eq.\ref{eq:cond}
is violated somewhere.
It is however straightforward to modify the formulae to account
for this.

\section{Preliminary numerical results}\label{sec:numer}
We have implemented the first option (single cluster)
to simulate a two-di\-men\-sional $Z(3)$ model with random coupling. The
coupling $\Omega_{ij}$ is uniformly distributed in the  region
\begin{equation}
\left(\half < \Real{\Omega}<\oneahalf \right)\; \cap  \;
\left( | \Arg{\Omega} | < \pi/3 \right).
\end{equation}

In order  to count sweeps in a fair
way, we defined a {\em sweep} as consisting
 of an update on $\Omega$ followed by
$N_{hits}$ cluster updates,
with $N_{hits}$  adaptively chosen in such a way that
$ N_{hits} \times {<N_c>} \approx L^2$, where ${<N_c>}$
is the average cluster size and $L$ is the lattice size. In this way
the computer time for a sweep is essentially independent on
$\beta$.
The autocorrelation time \tint has been measured for the total
magnetization $<\zeta>$ using the
formula given in \cite{Wolff2}.
We report the correlation length $\xi$ for $<\Real{\zeta_i \bar\zeta_j}>$
projected at vanishing transverse momentum, together with \tint and
the average linear size of the clusters.
Data have been taken on a $128 \times 128$ lattice on several
runs of 5000 sweeps each (Tab.\ref{tab:autoc}) and on a $200 \times 200$
lattice with 2500 sweeps only (Tab.\ref{tab:autoc2}).
\begin{table}[b]
\caption{$128\times 128$}\label{tab:autoc}
\begin{center}
\begin{tabular}{|c|c|c|c|} \hline
$\beta$ & $\xi$ & \tint & $\surd<N_c> $\\ \hline
.560 & 2.3 &  1.0 & 4 \\
.580 & 3.0 &  1.0 & 5 \\
.600 & 4.7 & 1.2 & 8 \\
.602 & 5.2 & 1.3 & 8 \\
.604 & 5.7 & 1.3 & 9 \\
.606 & 6.3 & 1.4 & 10\\
.608 & 7.7 & 1.9 & 11 \\
.610 & 12.0& 3.5 & 16 \\
.612 & $\approx$25 & $\approx$22 & 32 \\
\hline
\end{tabular}
\end{center}
\end{table}

\begin{table}[b]
\caption{$200\times 200$}\label{tab:autoc2}
\begin{center}
\begin{tabular}{|c|c|c|c|} \hline
$\beta$ & $\xi$ & \tint & $\surd<N_c> $\\ \hline
.600 & 4.7 & 1.0 & 7 \\
.602 & 4.9 & 1.2 & 8 \\
.604 & 5.5 & 1.3 & 9 \\
.606 & 6.2 & 1.5 & 10 \\
.608 & 7.6 & 1.6 & 11 \\
.610 & 9.3 & 2.5 & 14 \\
.612 & 18.0 & 12.0 & 23 \\
.613 & $\approx$25 & $\approx$21 & 33 \\
.614 & $\approx$32 & $\approx$40 & 41 \\
\hline
\end{tabular}
\end{center}
\end{table}
The numerical data show a clear sign of critical behaviour around
$\beta \approx 0.615 $.
At the same time the autocorrelation time \tint does not increase
more than linearly in $\xi$, at least up to $\beta \approx .610$;
beyond this value $\xi$ becomes a substantial
fraction of the lattice size and \tint starts to grow more
rapidly. To determine the dynamic critical exponent for the
algorithm one needs a very high statistics, and our data are
too preliminary to get a reliable conclusion. The sharp rise
in \tint near the transition, if confirmed, would mean that
the algorithm is not as efficient as we would have expected,
but it may still be of some practical value -- since the actual
values are not dramatically high. One also remarks that going
to bigger lattices tends to improve the performance.

\section{Conclusions}\label{sec:concl}
We have examined the feasibility of a cluster algorithm designed
to embed $Z(3)$ into $SU(3)$ in the spirit of Brower and Tamayo.
Clearly most of the
work is still to be done. We have to combine the present $Z(3)$
algorithm with some local update mechanism for the $SU(3)/Z(3)$
degrees of freedom. This should carefully deal with the
{\em coset} condition \ref{eq:coset} -- a heat bath technique
would probably be best suited. For the application to
gauge theory one has to further modify the algorithm, but
some progress was done recently on gauge $Z(2)$ and $U(1)$ systems,
and also this problem will hopefully be overcome.

What will the overhead be for
such kind of algorithm, say in the study of the deconfining transition?
Judging form recent progress in implementing
 cluster algorithms on massively
parallel machines\cite{FlanTam}\cite{Rossi}, the price to be paid
should not be too high.

\noindent
{\bf Acknowledgments}

\noindent
I would like to thank warmly W.\ A.\ Bardeen, P.\ B.\ Mackenzie,
R.\ K.\ Ellis, A.\ S.\ Kronfeld and the Theory Division
of Fermilab for their kind hospitality while this work was done.

\end{document}